# Cooling Rate Effects on the Structure of 45S5 Bioglass: Computational and Experimental Evidence of Si–P Avoidance

Pratik Bhaskar[1,#], Yashasvi Maurya[1,#], Rajesh Kumar[1], R. Ravinder[1], Amarnath R. Allu[2], Sumanta Das[3], Nitya Nand Gosvami[4], Randall E. Youngman[5], Mikkel S. Bødker[6], Nerea Mascaraque[6], Morten M. Smedskjaer[6], Mathieu Bauchy[7,*], N. M. Anoop Krishnan[1,4,*]

[1]Department of Civil Engineering, Indian Institute of Technology Delhi, Hauz Khas, New Delhi 110016, India
[2]Glass Division, CSIR-Central Glass and Ceramic Research Institute, 700032, Kolkata, India
[3]Department of Civil and Environmental Engineering, University of Rhode Island, Kingston, RI, USA
[4] Department of Materials Science and Engineering, Indian Institute of Technology Delhi, Hauz Khas, New Delhi 110016, India
[5] Science and Technology Division, Corning Incorporated, Corning, NY 14831, USA
[6] Department of Chemistry and Bioscience, Aalborg University, 9220 Aalborg, Denmark
[7] Physics of AmoRphous and Inorganic Solids Laboratory (PARISlab), Department of Civil and Environmental Engineering, University of California, Los Angeles, CA 90095, USA

[*]Corresponding author: M. Bauchy (bauchy@ucla.edu), N. M. A. Krishnan (krishnan@iitd.ac.in)
[#]Both the authors contributed equally.

**Abstract**
Due to its ability to bond with living tissues upon dissolution, 45S5 bioglass and related compositions are promising materials for the replacement, regeneration, and repair of hard tissues in the human body. However, the details of their atomic structure remain unclear. This is partially due to the non-equilibrium nature of glasses, as their non-crystalline structure is highly dependent on their thermal history, namely, the cooling rate used during quenching. Herein, using molecular dynamics (MD) simulations and magic angle spinning nuclear magnetic resonance (MAS-NMR) spectroscopy experiments, we investigate the structure of the nominal 45S5 bioglass composition prepared using cooling rates ranging over several orders of magnitude. We show that the simulations results are in very good agreement with experimental data, provided that they are extrapolated toward lower cooling rates achieved in experiments. These results highlight that previously reported inconsistencies between simulations and experiments stem from the difference in cooling rate, thereby addressing one of the longstanding questions on the structure of bioglass. Based on these results, we demonstrate the existence of a Si–P avoidance behavior, which may be key in controlling the bioactivity of 45S5 bioglass.

**Introduction**
Since its introduction by Hench, 45S5 bioglass remains one of the most promising bioactive glass compositions.[1] 45S5 bioglass, henceforth referred to as bioglass, is constituted by the oxides of calcium, sodium, phosphorus, and silicon and has the capacity to form strong interfacial bonds with hard tissues *in vivo* through dissolution of ions when they are hydrated by body fluids.[1] Bioactive glass compositions have been widely used for a variety of applications ranging from scaffolds for tissue engineering,[2] bone grafts,[3] periodontal applications,[4,5] and even as a coating for bioinert implants.[6,7] Despite the wide usage of bioactive glasses, some open questions regarding the atomic-level structure and other structural features of bioglass remain, which largely arise from some inconsistencies between previously reported simulation and experiment results.[8–10] This is exemplified by the fact that, although experimental studies propose the existence of isolated tetrahedra,[8–13] atomistic simulations of bioglass suggest the existence of phosphate tetrahedra with higher degrees of crosslinking. A detailed understanding of the structure of bioglass is essential to tailor the dissolution processes governing bioactivity.[10,11,14–16]

Bioglass samples, like other glasses, are out-of-equilibrium systems formed by the rapid quenching of an isochemical equilibrium liquid.[17,18] Accordingly, the structure and properties of bioglass strongly depend on the thermal and pressure history they have experienced during melt-quenching or subsequent processing. The structure of bioglasses is typically characterized using techniques such as infrared spectroscopy, Raman spectroscopy, magic angle spinning nuclear magnetic resonance (MAS-NMR) spectroscopy,[10–12] or using computational techniques such as molecular dynamics (MD) or density function theory (DFT) simulations.[8,9,19] MD simulations are widely used as they provide some insights into the atomic details that are otherwise invisible to conventional experiments. In particular, classical MD simulations are usually preferred over DFT simulations due to the latter's limitation to small sizes (typically < 500 atoms) and timescales (typically < 10 picosecond). Although classical MD simulations are also restricted to short timescales (typically a few nanoseconds), this timescale can be four to five orders of magnitude higher than that of DFT simulations. In turn, these short timescales (both for MD and DFT) lead to unrealistically high cooling rates in simulations, such as $10^{12}$ K/s, which is almost 10 orders of magnitude higher than typical experimental cooling rates (1-to-100 K/s). Therefore, understanding the effect of the cooling rate on the structure of bioglass provides the key for establishing a realistic, unified description of the atomic structure of bioglass.

Several earlier studies have been conducted to study the effect of cooling rates on model glasses,[20–22] vitreous silica,[23–25] and sodium silicate glass[26] structures. The high cooling rate used in simulations typically results in more structural defects.[26–28] Further, the cooling rate is found to have a notable impact on some of the macroscopic properties such as density, thermal expansion, and elastic modulus[26,29]—wherein the extent of the dependence on the cooling rate depends on the glass composition. Moreover, earlier studies based on MD simulations have suggested that, although the short range order is usually weakly affected, the medium range order can present notable changes with varying cooling rates.[26] This is important as, in the case of bioglass, the medium range order is suggested to play a major role in controlling the bioactivity.[11] This raises some concerns on the validity of the structure–property relations generated by MD simulations.

Herein, using MD simulations, we investigate the structure of bioglass samples prepared using cooling rates ranging over five orders of magnitudes. In line with previous results,[26] we find that the cooling rate strongly affects the medium-range order structure, whereas the short-range order remains fairly unaffected. Interestingly, in contrast to previous reports, we observe of an increasing propensity for Si–O–P avoidance upon decreasing cooling rate. Although the governing mechanisms may be different, this observation echoes the Loewenstein avoidance principle observed in aluminosilicates,[30] wherein Si and Al tend to preferentially form some asymmetric bonds at the expense of Al–O–Al symmetric linkages. Finally, by extrapolating the structural features obtained from MD simulations such as $Q^n$ distribution to experimental cooling rates, we demonstrate that MD simulations can be reconciled with our NMR measurements of $Q^n$ speciation to provide a consistent structure for bioactive glasses.

**Methodology**
**Simulation details**
**(i)  Glass simulation**

45S5 bioglass samples, with the nominal molar composition $(CaO)_{26.9}.(Na_2O)_{24.4}.(SiO_2)_{46.1}.(P_2O_5)_{2.6}$, were prepared using classical MD simulations by applying the traditional "melt-quench" method.[26,31,32] All simulations were performed using the open-source LAMMPS (Large-scale Atomic/Molecular Massively Parallel Simulator) package.[33] In order to study the effect of the cooling rate, the glass samples were prepared with cooling rates varying over five orders of magnitudes, from 100 K/ps to 0.01 K/ps. The interactions between the atoms were simulated using the well-established Teter potential,[8,26] which has a Buckingham-like form. This potential has been extensively validated against experimental structure data for a wide-range of silicate and phosphate minerals and glasses,[26,29,34–36] including 45S5 bioglass (see Supplementary Information and Ref.[8]). A repulsive term, $V(r) = B/r^n + Dr$, was adopted for $r$ smaller than $r_0$ (variable for each of the atom pairs), to avoid the "Buckingham catastrophe" at high temperature. Here, $r_0$ is defined as the value of $r$ when the third derivative of potential energy approaches zero. At $r_0$, $B$ and $n$ were chosen to make potential energy and its derivatives continuous. The long range Coulombic interactions were calculated with the particle-particle particle-mesh (PPPM) solver algorithm with an accuracy of $10^{-5}$.

Initial configurations were built by randomly placing 5673 atoms in a cubical box such that there is no unrealistic overlap. The system was then equilibrated for 1 ns at 3000 K and zero pressure in the isothermal isobaric (*NPT*) ensemble. This was to ensure that the system reaches an equilibrium state, without any memory of the initial structure obtained by the random placement of atoms. After equilibration, glasses were formed by gradually cooling the systems from 3000 to 300 K linearly at five different cooling rates, namely, 100, 10, 1, 0.1, and 0.01 K/ps under zero pressure in the *NPT* ensemble. Finally, all the glass structures were relaxed for 100 ps at 300 K and zero pressure in the *NPT* ensemble. For statistical averaging, additional simulations were carried out in an *NVT* ensemble for 100 ps. All the properties referring to the "glassy state" are obtained by averaging over 100 frames extracted from this run.

**(ii)  Pair distribution function**

The pair distribution function (PDF) represents the ratio of the local density of atoms with respect to the global density of the atoms in terms of distance from an atom. In particular, the neutron pair distribution function is given by:

$$g_N(r) = \frac{1}{\sum c_i c_j b_i b_j} \sum c_i c_j b_i b_j g_{ij}(r), \qquad \text{Equation (1)}$$

where $c_i$ is the fraction of $i$ atoms ($i$ = Na, Ca, Si, P, or O), $b_i$ is the neutron scattering length of the species, and $g_{ij}$ are the partial PDFs. Using the partial PDF, the coordination number of each of the species can be obtained by computing the number of neighbors within the first coordination shell of the respective atoms. The cutoff distance was obtained from the first minimum of the respective partial PDFs.

**(iii)  Structure factor**

The partial structure factors were calculated from the Fourier transformation of the partial pair distribution functions $g_{ij}(r)$ as:

$$S_{ij}(Q) = 1 + \rho_0 \int 4\pi r^2 [g_{ij}(r) - 1](\sin(Q_r)/Q_r)(\sin(\pi r/R)/(\pi r/R))dr, \qquad \text{Equation (2)}$$

where $Q$ is the scattering vector, $\rho_0$ is the average atom number density, and $R$ is the maximum value of the integration in real space, which is set to half of the size of one side of the simulation cell. The total structure factor was calculated as:

$$S_N(Q) = \frac{1}{(\sum c_j c_i b_j b_i)^2} \sum c_i c_j b_i b_j S_{ij}(Q), \qquad \text{Equation (3)}$$

where $c_i$ and $c_j$ are the fractions of atoms and $b_i$ and $b_j$ are neutron scattering lengths, for elements $i$ and $j$, respectively. The partial pair distribution functions and structure factors were plotted by taking statistical averages over 100 frames at 300 K. The position of the first sharp diffraction peak (FSDP) and the full width at half maxima (FWHM) of the structure factors were calculated by fitting the FSDP with a Lorentzian function for all the cooling rates.

### (iv)   $Q^n$ distribution

The $Q^n$ notation quantifies the degree of connectivity of the tetrahedra present in oxide glass networks. In this $Q^n$ notation, $n$ represents the number of oxygens that form a bridge between two tetrahedra—namely, bridging oxygen (BO)—around a network-forming cation. Thus, $Q^4$ represents a tetrahedron connected to four other tetrahedra through BO atoms. For 45S5 bioglass, the tetrahedra can be formed by $SiO_4$ and $PO_4$ species. Here, the $Q^n$ distribution was computed through an in-house code, which counts the number of BO associated with each Si/P atoms.

### (v)   Ring distribution

In glassy networks, a ring represents a closed path made of interatomic bonds. In particular, we focused on hetero-polar primitive rings that represent the shortest path made of Si–O and P–O bonds to return to the starting atom without retracing any route. The open-source RINGS package was used to compute the ring size distribution.[37] Note that in oxide glasses, the ring size is represented by the number of network formers in the ring, in this case, Si or P. Thus, a ring size of *m* contains *2m* atoms, namely, with *m* O atoms and *m* Si or P atoms.

**Experimental details**
### (i)   Glass preparation

The 45S5 glass was prepared using the melt-quenching method by mixing a batch of the following reagent grade materials: $Na_2CO_3$ (Ph Eur. ⩾99.5%, Sigma-Aldrich), $CaCO_3$ (Reag. Ph Eur, Merck), $Ca(H_2PO_4)_2 \cdot H_2O$ (Budenheim Ibérica S.L.U., ⩾95%), and $SiO_2$ (Sigma-Aldrich, purum p.a.). This batch, with its stoichiometric amounts, was melted in a Pt-Rh crucible at 1350 °C for 1 h in an electrically heated furnace. The melt was cast onto brass plates and the transparent glass was annealed for 30 minutes at its glass transition temperature ($T_g$) of 523°C. X-ray diffraction analysis shows no signs of crystallization.

### (ii)   NMR experiments

$^{29}Si$ MAS NMR was conducted at 11.7 T (99.27 MHz resonance frequency) using an Agilent DD2 spectrometer and 5 mm MAS NMR probe. The powdered bioglass was packed into a 5 mm outer diameter zirconia MAS NMR rotor, providing sample spinning of 5 kHz. $^{29}Si$ MAS NMR data were collected using radio-frequency pulse widths of 4.8 μs, corresponding to a tip angle of π/2, and with a 10 s recycle

delay, 3000 acquisitions were collected. The $^{29}$Si $T_1$ value was estimated to be on the order of 1-2 s, resulting from the incorporation of iron (as batch material impurity) in the glass sample. A $^{29}$Si MAS NMR spectrum collected with a much longer recycle delay (60s) was identical, confirming adequate relaxation time for the NMR experiments.

$^{31}$P MAS NMR data were collected at 16.4 T using an Agilent DD2 spectrometer, 3.2 mm MAS NMR probe (22 kHz sample spinning) and a resonance frequency of 283.27 MHz. 600 acquisitions were collected with a π/2 pulse width of 3 μs and a recycle delay of 5 s, optimized for the short $^{31}$P $T_1$. MAS NMR data were processed without apodization and shift referenced to tetramethylsilane and 85% phosphoric acid solution at 0.0 ppm for $^{29}$Si and $^{31}$P, respectively.

### (iii) Deconvolution of NMR results

Spectral deconvolution was made using DMFit[38]. The phosphate site ($Q^n$) concentrations were obtained from the simulations of the isotropic peaks and all associated spinning sidebands of the $^{31}$P MAS spectra. The $^{31}$P quantifications were used to constrain the silica sites, assuming charge neutrality in the glasses. The silica site concentrations were obtained from Gaussian simulations of the isotropic peaks and all of their associated spinning sidebands.

**Results:**

### (i) Enthalpy and density

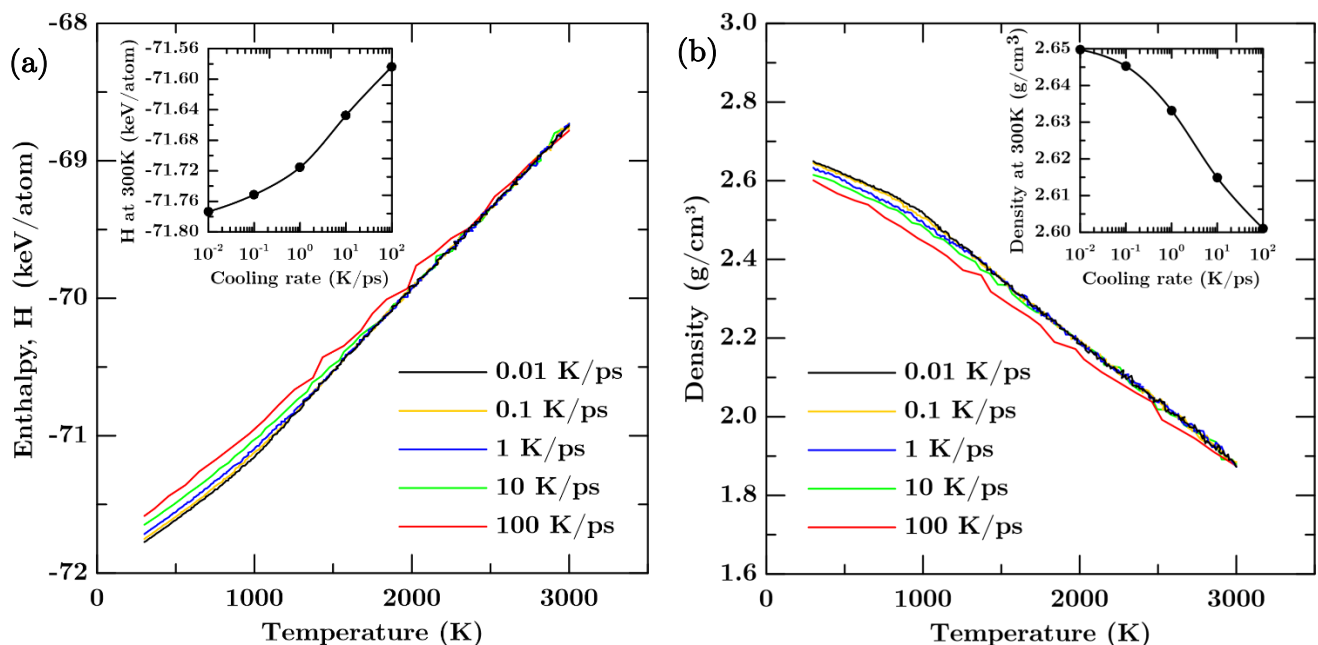

**Figure 1 (a).** Computed values of enthalpy (*H*) as a function of temperature during cooling under different rates by MD simulations. The inset shows *H* at 300 K as a function of the cooling rate. The line is a guide for the eye. **(b)** Computed values of density as a function of temperature during cooling under different rates by MD simulations. The inset shows density at 300 K as a function of the cooling rate. The line is a guide for the eye.

To demonstrate assess the effect of the cooling rate on the features of the glass transition, we first focus on the evolution of enthalpy ($H$) and density of the glass upon quenching computed by MD simulations. Figure 1(a) shows the evolution of the enthalpy with respect to temperature for five different cooling rates, namely, 0.01, 0.1, 1, 10, and 100 K/ps. Upon quenching, we observe that the enthalpy decreases continuously and monotonically with temperature. In particular, we observe that there is no sudden change in the enthalpy of the system, confirming the absence of any first-order phase transition. Further, the enthalpy exhibits a smooth and continuous change in slope at a particular temperature, namely the "fictive temperature" of the glass[39]. Note that the fictive temperature decreases with decreasing cooling rate, in agreement with the fundamental nature of the glass transition[26,40]. The inset of Fig. 1(a) shows the enthalpy at 300 K of the glasses obtained from different cooling rates. We observe that this enthalpy value decreases with decreasing cooling rate. This suggests that the glasses prepared with lower cooling rates are energetically more stable in comparison to those prepared at faster cooling rates. This can be attributed to the fact that, upon slower cooling rates, the system can sample a larger region of the energy landscape, thereby allowing it to attain lower, more stable energy states[40–43].

Figure 1(b) shows the evolution of density with respect to temperature computed by MD simulations for the five cooling rates considered. Similar to the case of the enthalpy, we observe a continuous and monotonic increase in the density of the glass with decreasing temperature—devoid of any first-order phase transition. Although a change in slope corresponding to the glass transition is also observed for the density, it does not occur at the same temperature as that of the enthalpy for each cooling rate. This is in agreement with earlier studies suggesting that the density and enthalpy relaxations in glassy systems at lower temperature may be decoupled[44,45]. The inset of Fig. 1(b) shows the cooling rate dependence of density of the glass at 300 K, which increases monotonically from 2.60 g/cm$^3$ for 100 K/ps to 2.65 g/cm$^3$ 0.01 K/ps, which suggest an enhanced degree of packing of the atoms in the system with slower cooling rates. Further, the simulated density of the glass tends towards the experimental value of 2.70 g/cm$^3$ with decreasing cooling rates[8]. Overall, these results confirm that our simulations offer a realistic description of the typical features of the glass transition.

### (ii) Short-range order and local structure

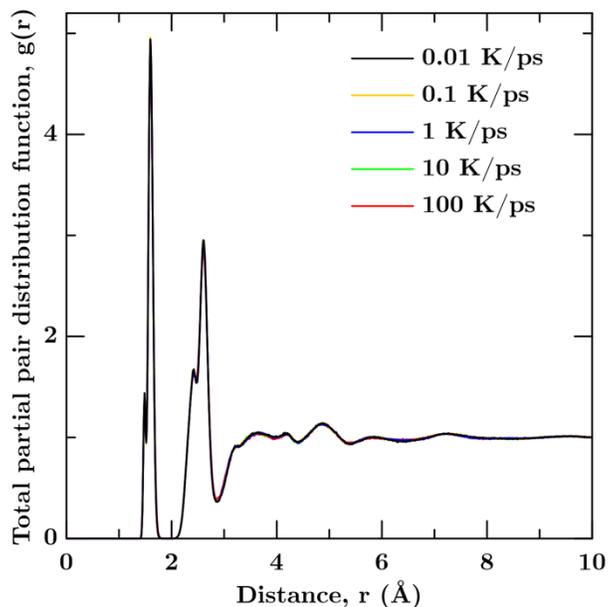

**Figure 2.** Total pair distribution functions in the glassy state for the studied cooling rates computed by MD simulations.

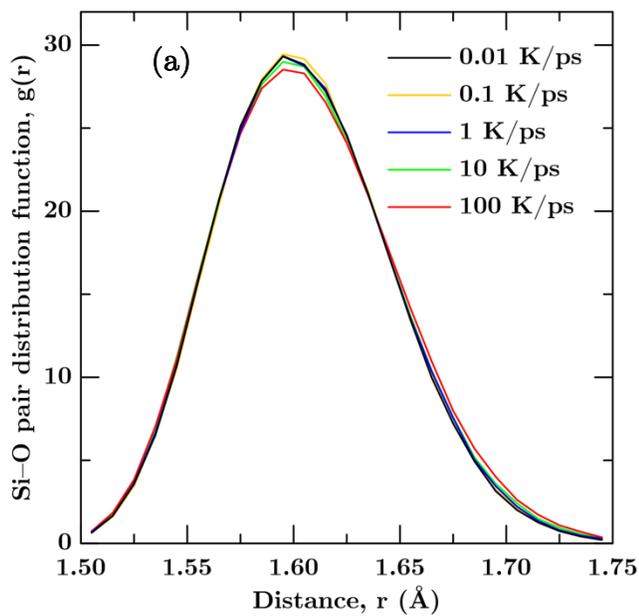
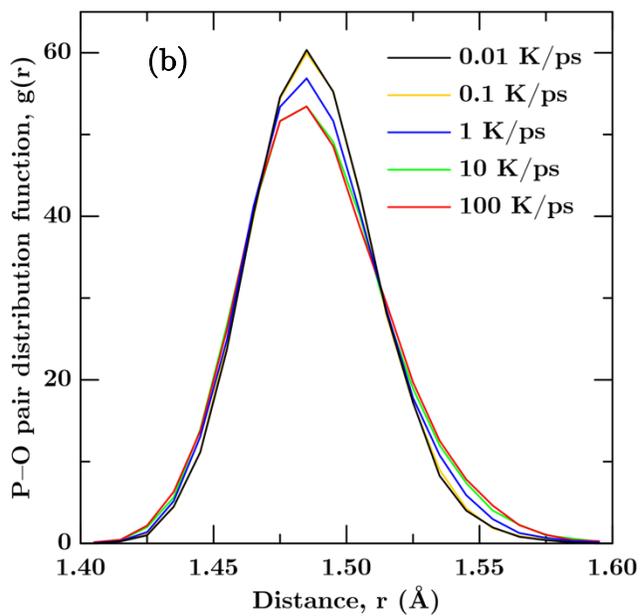

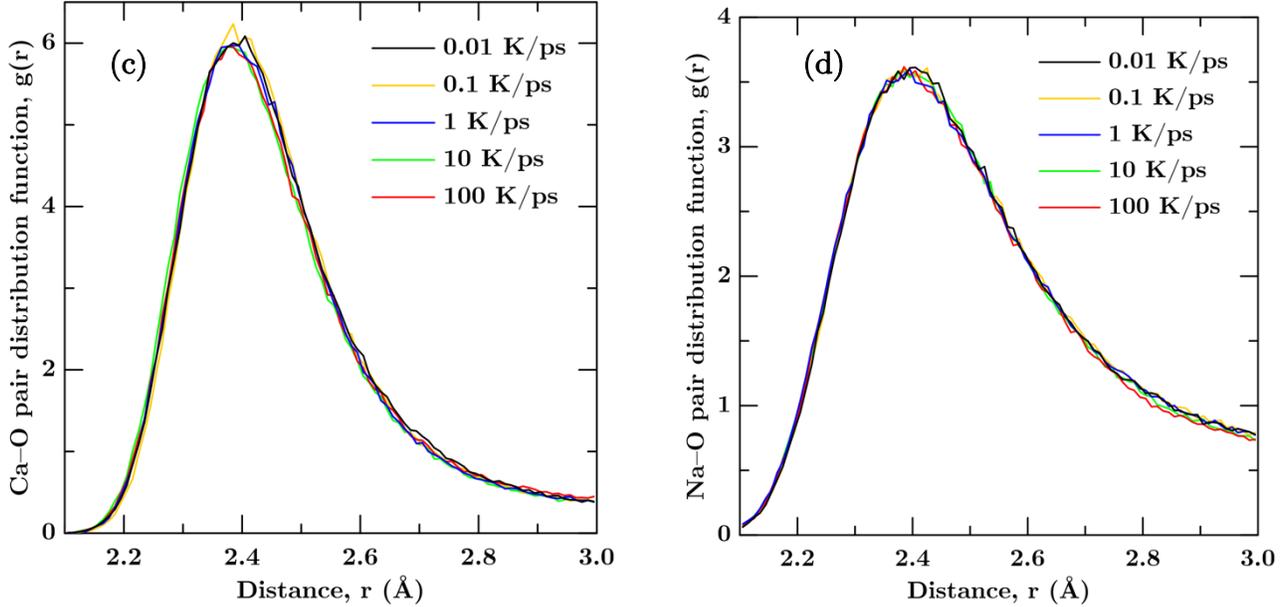

**Figure 3.** Hetero-nuclear partial pair distribution functions in the glassy state computed by MD simulations for the studied cooling rates of: **(a)** Si–O, **(b)** P–O, **(c)** Ca–O, and **(d)** Na–O.

Now, we analyze the short-range structure (≤ 3 Å) of the glasses obtained from MD simulations. Figure 2 shows the total pair distribution function (PDF) of the glassy structures for different cooling rates. The first peak observed around 1.5 Å corresponds to the P–O interactions, which is followed by a major peak observed around 1.6 Å corresponding to the Si–O interactions. Note that the values obtained here are in agreement with previous simulations, and experiments (See Supplementary Information and Refs. [8–10]). The following peaks up to 3 Å correspond to Na–O, Ca–O, and O–O interactions. Beyond 3 Å, we observe minor peaks corresponding to second neighbor and third neighbor interactions. At larger distances, the PDF gradually converges to unity, confirming the absence of any long-range order in the system. Further, we observe that the position of the peaks of the total PDF varies only slightly, if any, with the cooling rate. This suggests that the first neighbor environments in the atomic structure are not significantly affected by the cooling rates. To further confirm this, we plot the X–O partial PDFs (with X = Si, P, Na, and Ca). In Fig. 3, we observe that for the X–O partial PDFs, neither the position nor the width of the peaks, are affected significantly by the cooling rate.

Next, we analyze the effect of cooling rate on the homo-nuclear interactions. Figure 4 shows the X–X partial PDFs with X = Si, O, Na, and Ca. We observe that the X–X interactions peak generally become sharper with decreasing cooling rate. This suggests that the overall degree of order among increases with decreasing cooling rate. Although this effect is pronounced in the Si–Si partial PDF (see Fig. 4(a)), it is less pronounced in the other pairs (see Fig. 4(b)–(d)). The Na–Na and Ca-Ca partial PDFs exhibit broader peaks as seen in Figures 4(c) and (d), respectively. This is due to the increased mobility of these atoms, thanks to the non-directional ionic bonds they form with oxygen. Other partial PDFs also exhibit similar behavior wherein the first peak becomes increasingly sharper corresponding to a decreasing cooling rate (see Figs. S2–S5 of the Supplementary Material) suggesting an increased local order. This is in agreement with the earlier observation that the structure attains a lower energy stable state as the cooling rate decreases (see Fig. 1(a)).

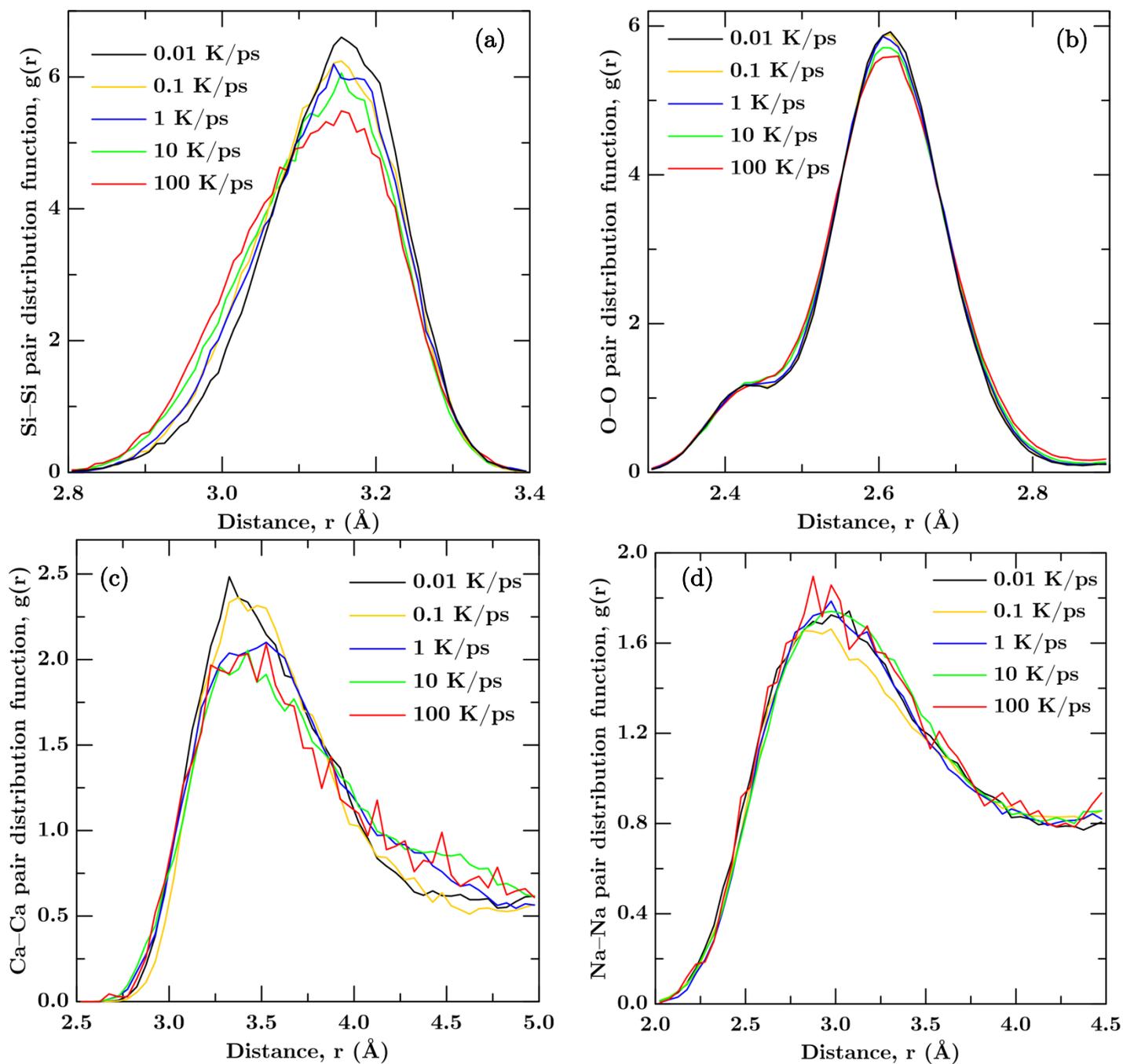

**Figure 4.** Homo-nuclear partial pair distribution functions in the glassy state, computed by MD simulations for the studied cooling rates, of: **(a)** Si–Si, **(b)** O–O, **(c)** Ca–Ca, and **(d)** Na–Na.

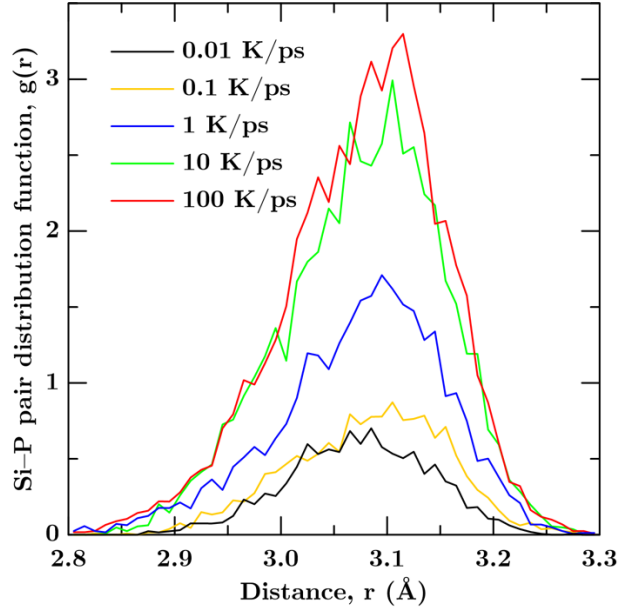

**Figure 5.** Si–P partial pair distribution functions in the glassy state for selected cooling rates computed by MD simulations.

Interestingly, we observe that the Si–P partial PDF exhibits a unique behavior as a function of cooling rate (see Fig. 5). Upon decreasing cooling rate, the intensity of the Si–P peak diminishes and the peak becomes notably less intense. Note that this is in contrast to the other partial PDFs (Figs. 3 and 4), which exhibit a sharper peak with decreasing cooling rate. This suggests that there is a preferential mutual avoidance between Si and P atoms, which is magnified upon decreasing cooling rate, i.e., when the glass reaches a more stable state. In other words, for a system obtained with reasonably low cooling rates, such as the experimental glasses, there is a tendency to avoid the formation of Si–O–P bridges. This observation is consistent with experimental studies (as demonstrated later), which report a low fraction of Si–O–P bridges, in agreement with earlier studies.[10,11] In order to demonstrate this further, we compute the fraction of Si–O–P and Si–O–Si linkages and compare them with the predictions from a random network model.[46] According to this model, each network former (that is, Si or P) can pick its neighbor with equal probability from the available atoms, irrespective of the atom type. Thus, the fractions of Si–O–Si, Si–O–P, P–O–P, angles are given by

$$f_{Si\text{-}O\text{-}Si} = \frac{4N_{Si} \times 4N_{Si}}{(4N_{Si} + 4N_{Si})^2}$$

$$f_{Si\text{-}O\text{-}Si} = \frac{4N_{Si} \times 3N_P}{(4N_{Si} + 43)^2} \qquad \text{Equation (4)}$$

$$f_{P\text{-}O\text{-}P} = \frac{3N_P \times 3N_P}{(3N_P + 3N_P)^2}$$

where $N_{Si}$ and $N_P$ are the numbers of Si and P atoms, respectively.

Figure 6 shows the fraction of the Si–O–Si and Si–O–P linkages predicted from the random model compared with the results from MD simulations for different cooling rates. We observe that, overall, the glasses exhibit a chemically-ordered structure that deviates from the predictions of the random model—wherein the computed fractions of Si–O–P and Si–O–Si linkages are lower and higher than those obtained from a random distribution, respectively. Note that the P–O–P linkages are absent in structures generated from MD simulations. As the cooling decreases, the deviation from the random increases and the glass exhibits an enhanced chemically-ordered nature. Note that this behavior echoes the Loewenstein exclusion principle[30] observed in aluminosilicate (borosilicate) glasses and minerals, where asymmetric Si–O–Al (Si–O–B) linkages are favored at the expense of symmetric Al–O–Al (B–O–B) bridges.[30,46] However, the Loewenstein rule essentially results from the preferential repulsion of charged $AlO_4$ polytope, whereas, here, the Si–O–P avoidance seems to have an energetical origin (since entropy would otherwise favor a more random structure). Overall, these results confirm that the Si–O–P bridges in bioglass exhibit an avoidance principle—the manifestation of which is closely linked to the cooling rate used to prepare the glass.

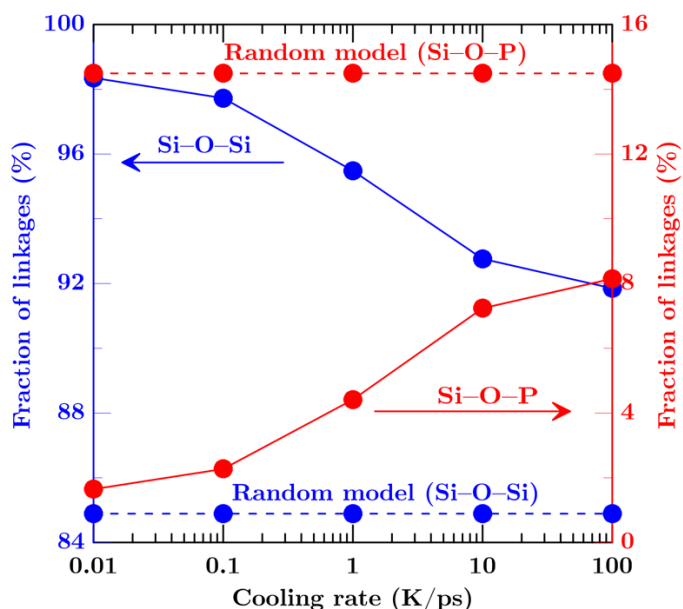

**Figure 6.** Si–P partial pair distribution functions in the glassy state for selected cooling rates.

Figures 7(a) and (b) show the intra-tetrahedral O–P–O and O–Si–O bond angle distributions (BADs), respectively. We observe that the average intra-tetrahedral angles (i.e., around 109°) are not significantly affected by the cooling rate, although the peaks become slightly sharper for lower cooling rates. This suggests an increased degree of angular order upon lower cooling rate. In the case of O–P–O, the average bond angle decreases slightly with decreasing cooling rate up to 1 K/ps. Below 1 K/ps the value exhibits little change. In the case of O–Si–O, we observe that the average intra-tetrahedral bond angle is not affected by the cooling rate. This suggests that the overall tetrahedral structure in the studied bioglass is not significantly affected or distorted by high cooling rates.

Figure 8 shows the inter-tetrahedral BADs. Similar to the intra-tetrahedral BADs, we observe that the Si–O–Si BAD exhibits little variation with different cooling rates. Interestingly, we observe that the peak intensity of BAD for Si–O–P decreases and the distribution becomes broader with decreasing cooling rates. As in the case of Si–P PDF (Fig. 5), this suggests an increased degree of angular disorder within the Si–O–P bridges with decreasing cooling rate, which arises from the propensity for Si–P mutual avoidance.

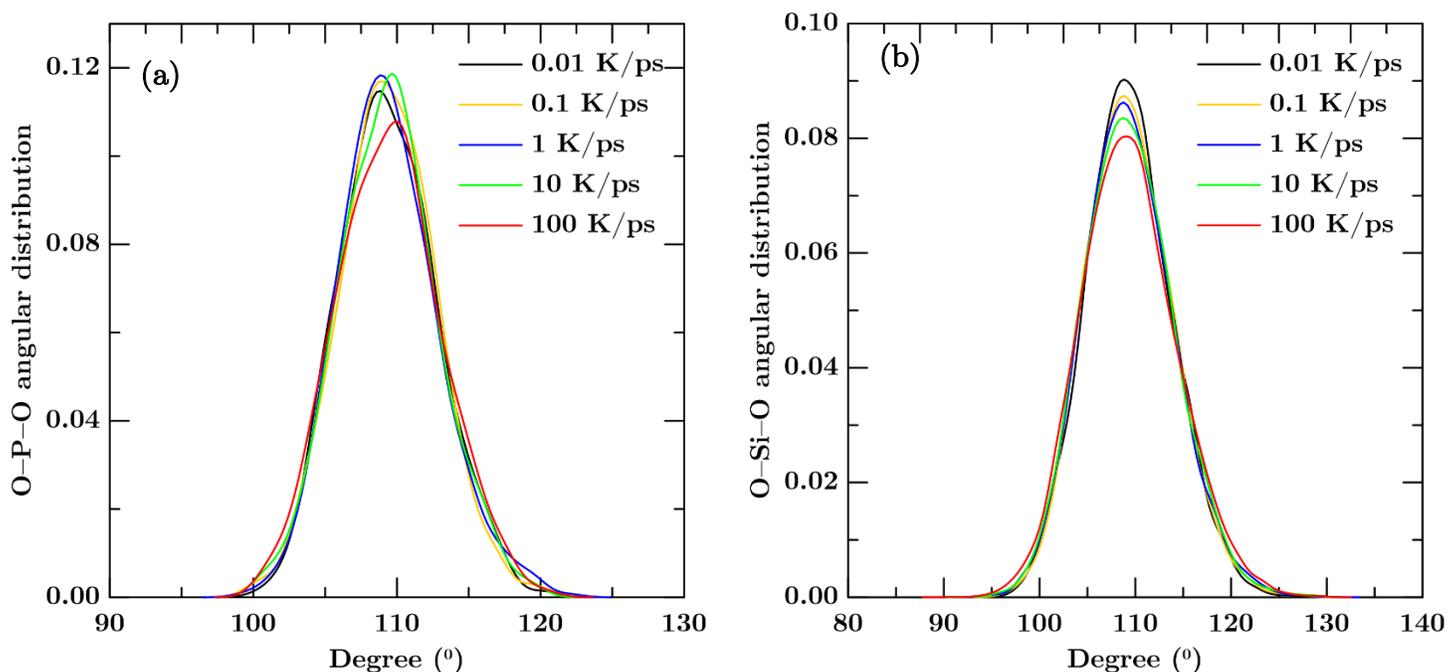

**Figure 7.** Intra-tetrahedral (a) O–P–O bond and **(b)** O–Si–O bond angle distribution (BAD) in the glassy state computed by MD simulations for selected cooling rates.

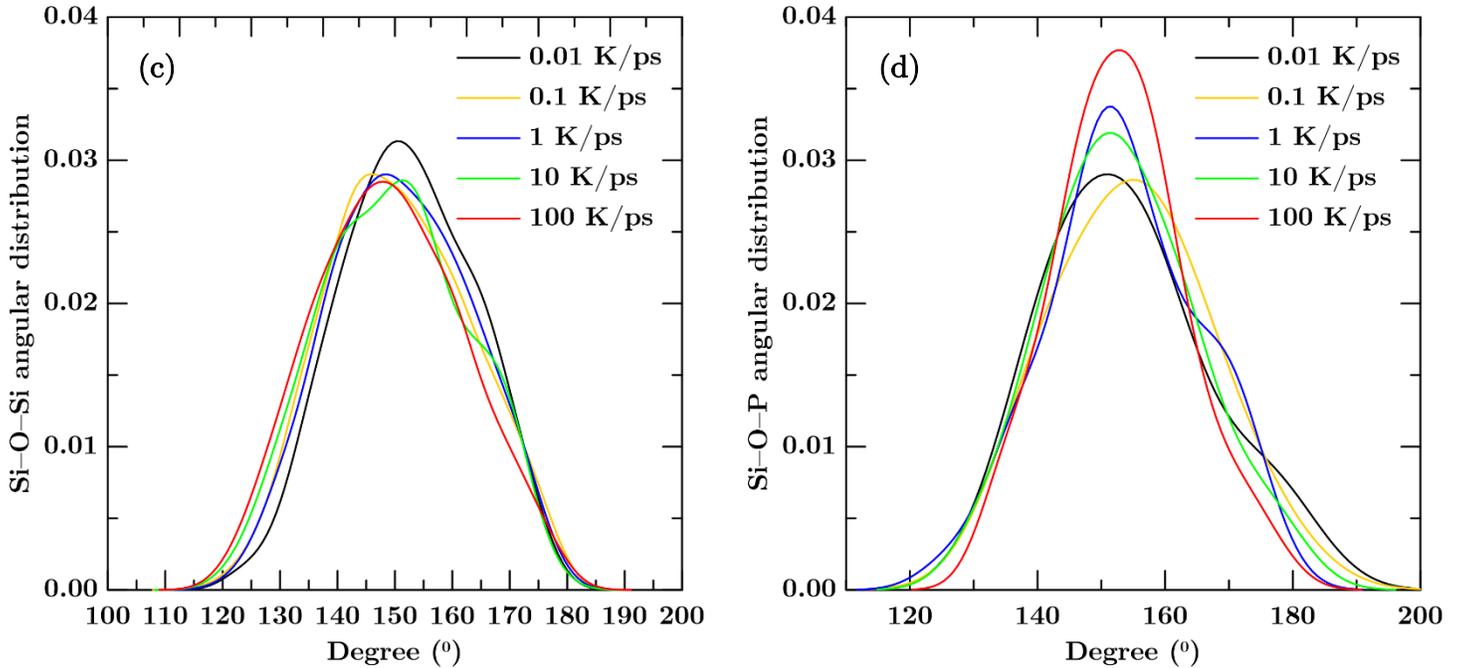

**Figure 8.** Inter-tetrahedral **(a)** Si–O–Si and **(b)** Si–O–P bond angle distribution (BAD) in the glassy state computed by MD simulations for selected cooling rates.

(iii) **Network connectivity**

To analyze the effect of the cooling rate on the network connectivity, we quantify the topology of the oxygen atoms. In oxide glasses, the network formers such as Si or P atoms form tetrahedra, which are connected by some BO atoms through X–O–Y bridges (X or Y = Si or P), thereby increasing the connectivity (or rigidity) of the network. In contrast, network modifiers such as Na or Ca atoms tend to depolymerize the network by breaking the X–O–Y bridges leading to the formation of X–O–Na/Ca bonds. These oxygen atoms are termed as non-bridging oxygen (NBO) atoms. Note that the addition of an Na atom creates one NBO, while the addition of a Ca atom results in two NBOs[17]. Figure 9 shows the fraction of BO and NBO atoms with respect to the cooling rate. Interestingly, we observe that the fraction of BO atoms decreases with decreasing cooling rate, while that of NBO increases. This can be attributed to the Si–O–P mutual avoidance (see Fig. 5 and 6) with decreasing cooling rates, leading to the increased depolymerization of the network, which results in the formation of NBOs. Overall, we observe that the degree of connectivity of the network decreases with decreasing cooling rate.

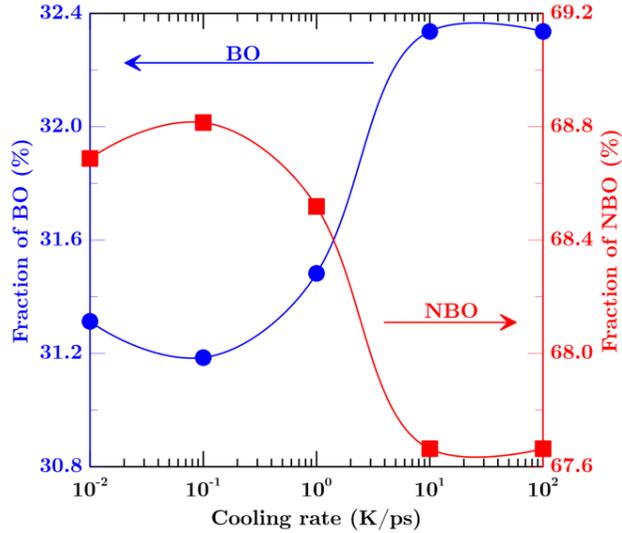

**Figure 9.** Fractions of Bonding Oxygen (BO) on the left y-axis and Non-Bonding Oxygen (NBO) on the right y-axis as a function of cooling rate. The line is a guide for the eye.

### (iv) $Q^n$ distribution

To study the connectivity of the tetrahedral network further, we analyze the $Q^n$ distribution (see Methodology section) both computationally and from solid-state NMR measurements on a bioglass sample. The fitting parameters used herein for both the $^{31}$P and $^{29}$Si MAS NMR data, as well as the obtained $Q^n$ distributions, are presented in Tables 1 and 2.

**Table 1.** Parameters used for deconvolution of $^{31}$P MAS NMR spectra of 45S5 bioglass. The $Q^n$ populations (atom %) include intensity from the spinning sidebands.

| $Q^0$ (%) | Pos (ppm) | Width (ppm) | Gaus/Lor (%) | | $Q^1$ (%) | Pos (ppm) | Width (ppm) | Gaus/Lor (%) |
|---|---|---|---|---|---|---|---|---|
| **97.38** | 8.91 | 8.03 | 0.78 | | **2.62** | 1.79 | 8.08 | 1 |

**Table 2.** Parameters used for deconvolution of $^{29}$Si MAS NMR spectra of 45S5 bioglass. The $Q^n$ populations (atom%) also include all spinning sideband intensity.

| $Q^1$ (%) | Pos (ppm) | Width (ppm) | | $Q^2$ (%) | Pos (ppm) | Width (ppm) | | $Q^3$ (%) | Pos (ppm) | Width (ppm) |
|---|---|---|---|---|---|---|---|---|---|---|
| **5.55** | -69.81 | 7.83 | | **76.69** | -78.4 | 10.45 | | **17.75** | -86.33 | 11.8 |

Figures 10(a) and (b) shows the $Q^n$ distributions of the tetrahedral species in the system formed by Si and P atoms, respectively. In the case of Si, MD simulations results suggest that a large fraction (around 40%) of the glass network comprises $Q^2$ tetrahedra, at the expense of $Q^3$ or $Q^4$ tetrahedra. Further, we observe that the fraction of $Q^0$, $Q^1$, and $Q^4$ Si units decreases monotonically as the cooling rate decreases. These values are then compared to experimental results obtained using $^{29}$Si MAS-NMR, as shown in Figure 11(a). We observe that the fraction of $Q^2$ units is approximately 76%, while that of $Q^1$ and $Q^3$ are approximately 5% and 17%, respectively (Table 2). These results suggest that the silica network primarily consists of $Q^2$ tetrahedra, with some degree of cross-linking provided by $Q^3$ species. Overall, we observe that the trend predicted by MD simulations exhibits a good match with experiments for the $Q^n$ units of Si atoms.

In the case of the P atoms, we observe that the structure exhibits only $Q^0$, $Q^1$, and $Q^2$ units, even at the highest cooling rates (see Fig. 10(b)). Upon decreasing cooling rate, $Q^2$ units decrease quickly and converge to zero, even within the range of cooling rates accessible by MD simulations. Further, the fractions $Q^0$ and $Q^1$ units monotonically increase and decrease with decreasing cooling rate, respectively. This means that, as the cooling rate decreases, P atoms increasingly tend to form some isolated $PO_4^{-3}$ tetrahedral units. These values are then compared to the experimental results obtained using $^{31}P$ MAS-NMR, the spectrum of which is plotted in Figure 11(b). Interestingly, we observe that P atoms exists only as $Q^0$ and $Q^1$ units for the cooling rate used experimentally, wherein the fraction of $Q^0$ atoms is around 95% (Table 1). This confirms that the P atoms prefer to exist mostly as isolated $Q^0$ orthophosphate ions in the bioglass system. It should be noted that a direct extrapolation may lead to the underestimation of the fraction of $Q^1$ units. This could be due to the low amount of P atoms present in the bioglass system, making an accurate determination of the $Q^n$ distribution challenging. Overall, we observe that the effect of the cooling rate on the $Q^n$ distributions can be reasonably well captured when MD simulations are extrapolated to experimental timescales, and furthermore, the experimentally-determined network structure of bioglass is consistent with previous studies of bioglasses[47].

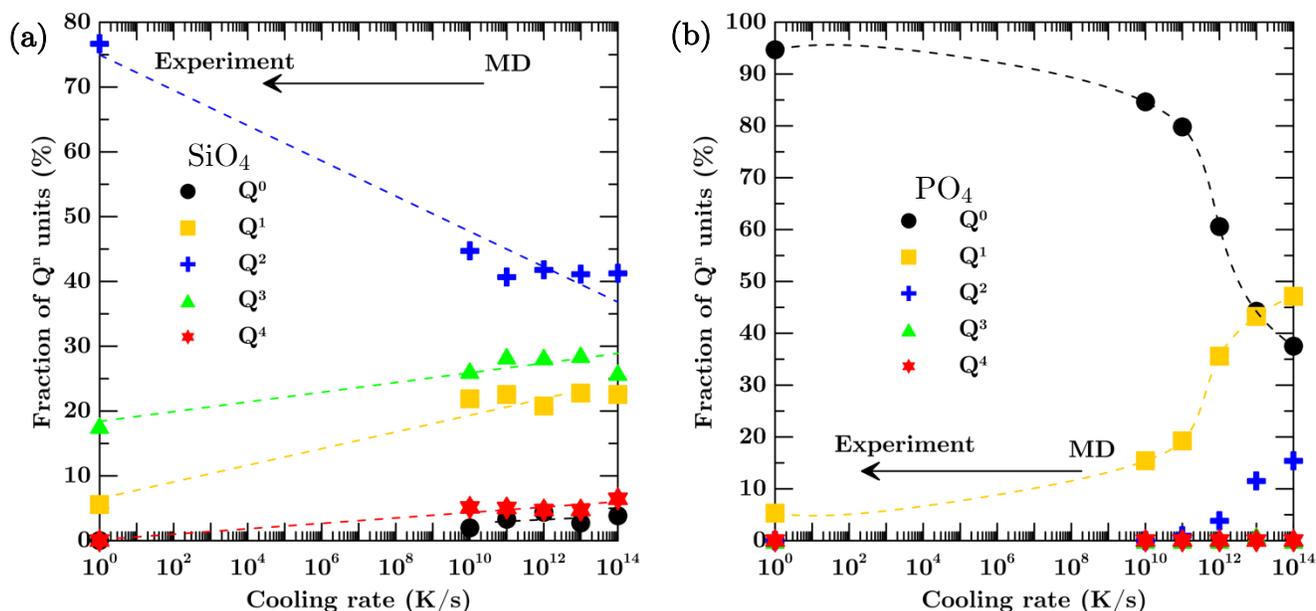

**Figure 10.** $Q^n$ distribution of **(a)** Si and **(b)** P tetrahedra in 45S5 Bioglass computed by MD simulations for five different cooling rates. Experimental data obtained by MAS-NMR are added for comparison. The lines are to guide the eye.

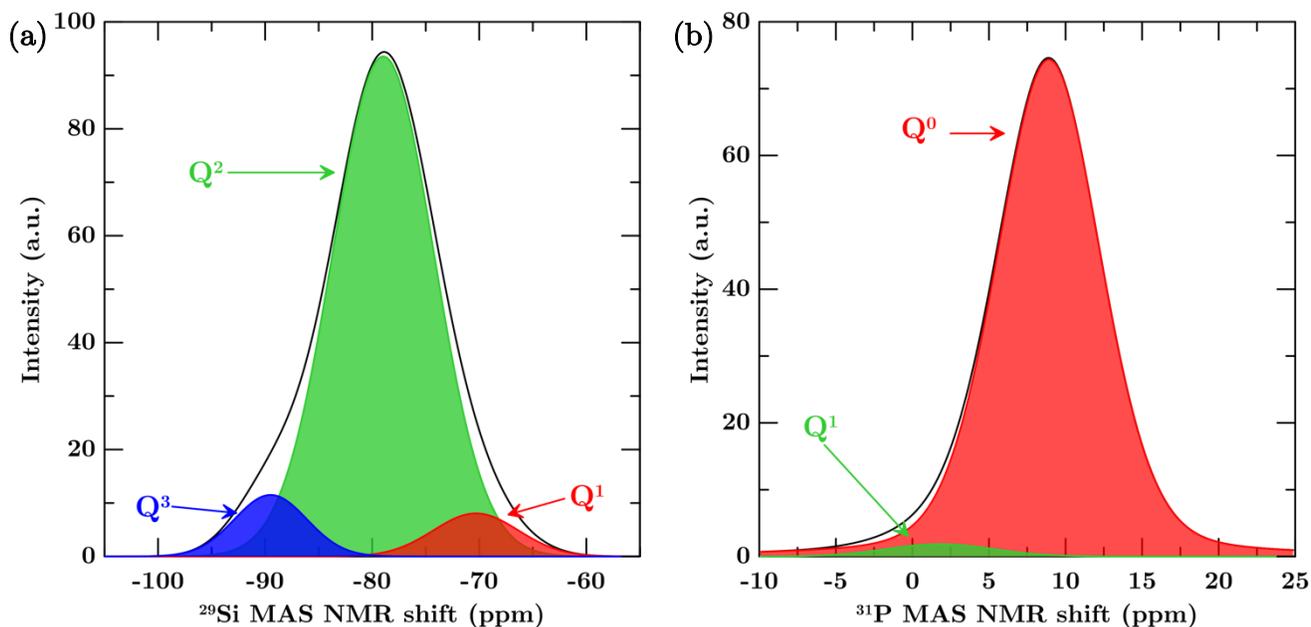

**Figure 11. (a)** $^{29}$Si MAS NMR spectrum of bioglass with deconvolution into $Q^1$, $Q^2$ and $Q^3$ silicate groups and **(b)** $^{31}$P MAS NMR spectrum of bioglass with deconvolution into $Q^0$ and $Q^1$ phosphate groups.

### (v) Medium-range order

While the PDF is useful for analyzing the structure correlations at smaller distances (≤ 3 Å), the structure factor represents the correlations at larger distances (between 3 Å to 10 Å). In particular, the first sharp diffraction peak (FSDP) captures the extent of correlation in the medium-range structure of glasses.[27,48–51] The full width at half-maximum (FWHM) of the FSDP is inversely linked to a medium-range coherence length $L$ as, $L = 7.7/\text{FWHM}$,[27,48–51] which captures the medium-range correlation distance among fluctuations of atomic density. Figure 12(a) shows the neutron structure factors computed by MD simulations with different cooling rates. We observe that the structure factor at larger values of wave vector exhibits little differences. However, FSDP becomes sharper with increasing peak intensity for decreasing cooling rates. This indicates that, in turn, the coherence length in the medium-range order increases upon decreasing cooling rate. To confirm this, we plot the FWHM and coherence length in Fig. 12(b). We observe that, upon decreasing cooling rate, FWHM decreases and the coherence length increases (see Fig. 12(b)), which denotes an increased degree of order within the medium-range structure of the glass. This echoes the fact that, due to the mutual avoidance between Si and P atoms (see Fig. 6), the network presents an increasingly chemically-ordered structure upon decreasing cooling rate—in agreement with the experimental and simulation data on $Q^n$ as well.

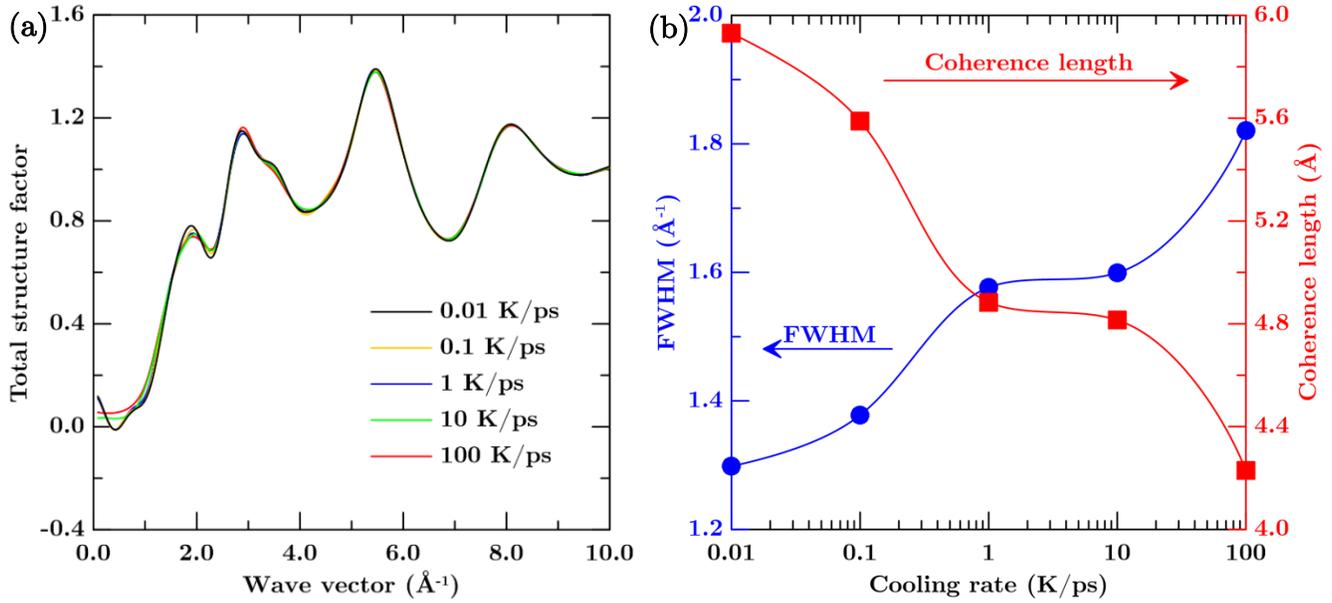

**Figure 12. (a)** Computed total neutron structure factors in the glassy state for selected cooling rates. **(b)** Full-width at half maximum (left axis), and coherence length (right axis) as a function of cooling rate. The line is a guide for the eye.

Finally, we analyze the effect of the cooling rate on the ring size distribution within the network computed by MD simulations (see Methodology section). As shown in Fig. 13, we find that, in contrast to the short-range order, the ring distribution is largely affected by the cooling rate. First, we observe that there are no two-membered rings, which confirms the absence of any edge-sharing tetrahedra. Further, there are no small three-membered rings for all but the highest cooling rate of 100 K/ps. This arises from the fact that, at high cooling rate, atoms may get frozen in unstable states due to the limited time available for structural rearrangements. With decreasing cooling rate, we observe that larger rings are preferred with a maximum fraction observed for five-membered rings, that is, rings having five Si atoms. In particular, a sharp peak at a ring size of 5 is observed for the cooling rate of 0.01 K/ps, which shows that almost 50% of the rings have a size of five Si atoms. This is in contrast with glassy silica, wherein a ring size of six is preferred.[27,28,52] Finally, we observe that all the rings are formed by Si and O atoms, thereby confirming that the P atoms do not participate in the ring formation. This is consistent with the fact that P atoms mostly form isolated $Q^0$ and terminating $Q^1$ units and, hence, are unable to participate in these ring structures. Overall, we conclude that the medium range structure of the bioglass is significantly affected by the cooling rate.

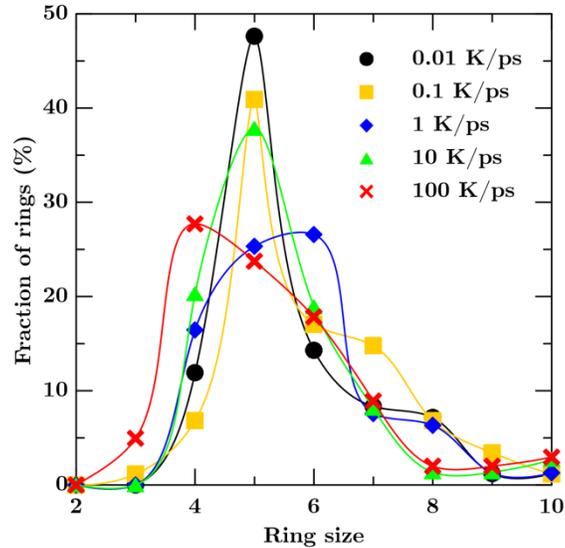

**Figure 13.** Ring distribution for the bioglass at various cooling rates computed by MD simulations.

**Discussions**

Earlier studies of bioglass using simulation and experiments have found some inconsistencies regarding structure of bioglass. Experimental studies based on $^{31}$P MAS-NMR spectroscopy revealed a broad resonance at around 9 ppm corresponding to isolated orthophosphate anions ($PO_4^{3-}$).[10,12,13] These isolated orthophosphate anions associate with metal cations, thereby reducing the network-modifying role of Ca/Na cations in the silicate network.[12] However, earlier MD simulations of bioglass contradicted these results by suggesting the existence of Si–O–P bridges with P existing as $Q^1$ and $Q^2$ species.[8,9] To reconcile these two results, a later experimental study suggested that, although it is possible that P may exist as $Q^1$ or $Q^2$ species, the levels of these species must be significantly lower than those suggested by MD simulations.[10]

The present results suggest a holistic picture reconciling earlier experimental and MD-simulation-based studies. While the short-range structure of bioglass hardly depends on the cooling rate, the medium range structure is significantly affected. In particular, our results confirm that Si and P atoms exhibit a mutual avoidance, thereby leading to the gradual disappearance of Si–O–P bridges at lower cooling rates. Further, lower cooling rate results in an increase in the concentration of isolated orthophosphate anions, which associate with the Ca and Na cations present in the structure. As such, the extrapolation of MD results toward lower experimental cooling rates would result in: (i) a significant decrease in the fraction of Si–O–P bridges, (ii) the absence of any $Q^2$ P species, as observed in our NMR measurements, and (iii) an increased propensity for P atoms to exist as isolated orthophosphate ions. Indeed, it is well known that the high cooling rates used in MD simulations tend to overestimate the degree of randomness in computed $Q^n$ distributions. However, it is interesting to note that the high cooling rates also results herein in a preferential formation of BOs (as in the case of Si–O–P) that are otherwise unfavorable at typical experimental cooling rates. Overall, the present study emphasizes the importance of using an appropriate cooling rate in MD simulations to reconcile experimental and simulation structural studies.

From a more general perspective, these results can have significant ramifications in the understanding of the structure and bioactivity of bioglass. It is well known that bioactivity is closely related to the leaching of ions from the glass during the dissolution process.[53] This occurs through the breakage of interatomic bonds present in the glass structure. As such, the absence of a large number of Si–O–P bridges, which were otherwise postulated by previous MD studies, suggests that isolated phosphate ions may exhibit an increased mobility and, hence, can easily leach out from the glass. This is further confirmed by the NMR experiments wherein approximately 95% of P atoms are shown to exist as $Q^0$ units. This may in turn contribute to the bioactive behavior of the glass. However, confirming this hypothesis would require future experimental studies on the degree of congruency in the leaching rates of Si and P species in bioglass.

**Conclusions**

The present study reveals the importance of the cooling rate in governing the structure of bioactive glasses. In particular, we show that the medium-range order structure is largely controlled by the thermal history of the glass. Interestingly, we observe a strong Si–P mutual avoidance leading to a decrease in the fraction of Si–O–P bridges upon lower cooling rates. This observation, along with the propensity for P atoms to form isolated tetrahedra, confirms the earlier experimental findings that phosphorus atoms primarily exist as isolated orthophosphate ions in bioglass. Further, we show from the ring distribution analysis that P does not take part in the formation of rings, thereby allowing increased mobility for P atoms even for $Q^1$ species. This structural feature can, in turn, be key to understand the bioactivity of this system as it could be related to the preferential leaching of P ions from the glass. Finally, through the present cooling rate analysis, our study reconciles the previous inconsistencies existing among experimental and simulation-based observations.

**Acknowledgement**

MMS acknowledges funding from the Independent Research Fund Denmark (grant no. 7017-00019). MB acknowledges funding from the National Science Foundation under Grants No. 1562066, 1762292, and 1826420. NMAK acknowledge the financial support for this research provided by the Department of Science and Technology, India under the INSPIRE faculty scheme (DST/INSPIRE/04/2016/002774) and DST SERB Early Career Award (ECR/2018/002228). Authors thank Corning Inc. for the experimental resources and IIT Delhi HPC facility for providing the computational and storage resources.

**Declaration of interests**

The authors declare that they have no known competing financial interests or personal relationships that could have appeared to influence the work reported in this paper.